\documentclass[12pt,manuscript,doublespace]{aastex}

\usepackage{natbib}                
\usepackage{verbatim}
\usepackage{subfigure}
\usepackage{epsfig}

\def\arcsec{\ifmmode '' \else $''$\fi}

\def\arcsecpoint{\ifmmode ''\!. \else $''\!.$\fi}

\def\kms{\ifmmode {\rm km\ s}^{-1} \else km s$^{-1}$\fi}
\def\Msun{\ifmmode {\rm M}_{\odot} \else M$_{\odot}$\fi}
\def\Lsun{\ifmmode {\rm L}_{\odot} \else L$_{\odot}$\fi}
\def\Zsun{\ifmmode {\rm Z}_{\odot} \else Z$_{\odot}$\fi}

\def\ergscm2{ergs\,s$^{-1}$\,cm$^{-2}$}
\def\icm3{{\rm cm}^{-3}}
\def\icm2{{\rm cm}^{-2}}
\def\qo{\ifmmode q_{\rm o} \else $q_{\rm o}$\fi}
\def\Ho{\ifmmode H_{\rm o} \else $H_{\rm o}$\fi}
\def\ho{\ifmmode h_{\rm o} \else $h_{\rm o}$\fi}

\def\vFWHM{\ifmmode v_{\mbox{\tiny FWHM}} \else
            $v_{\mbox{\tiny FWHM}}$\fi}
\def\CCF{\ifmmode F_{\it CCF} \else $F_{\it CCF}$\fi}
\def\ACF{\ifmmode F_{\it ACF} \else $F_{\it ACF}$\fi}
\def\Halpha{\ifmmode {\rm H}\alpha \else H$\alpha$\fi}
\def\Hbeta{\ifmmode {\rm H}\beta \else H$\beta$\fi}
\def\Hgamma{\ifmmode {\rm H}\gamma \else H$\gamma$\fi}
\def\Hdelta{\ifmmode {\rm H}\delta \else H$\delta$\fi}
\def\Lya{\ifmmode {\rm Ly}\alpha \else Ly$\alpha$\fi}
\def\Lyb{\ifmmode {\rm Ly}\beta \else Ly$\beta$\fi}
\def\Lyg{\ifmmode {\rm Ly}\gamma \else Ly$\gamma$\fi}

\def\hi{H\,{\sc i}}

\def\cii{C\,{\sc ii}}
\def\ciii{\ifmmode {\rm C}\,{\sc iii} \else C\,{\sc iii}\fi}
\def\civ{\ifmmode {\rm C}\,{\sc iv} \else C\,{\sc iv}\fi}
\def\cv{\ifmmode {\rm C}\,{\sc v} \else C\,{\sc v}\fi}
\def\cvi{\ifmmode {\rm C}\,{\sc vi} \else C\,{\sc vi}\fi}

\def\nv{N\,{\sc v}}

\def\o5007{[O\,{\sc iii}]\,$\lambda5007$}

\def\ovi{O\,{\sc vi}}

\def\siiv{Si\,{\sc iv}}

\def\siII{Si\,{\sc ii}}

\def\pv{P\,{\sc v}}

\def\o{\o}

\def\gtorder{\mathrel{\raise.3ex\hbox{$>$}\mkern-14mu
             \lower0.6ex\hbox{$\sim$}}}
\def\ltorder{\mathrel{\raise.3ex\hbox{$<$}\mkern-14mu
             \lower0.6ex\hbox{$\sim$}}}
\def\proptwid{\mathrel{\raise.3ex\hbox{$\propto$}\mkern-14mu
             \lower0.6ex\hbox{$\sim$}}}

\begin{document}

\shortauthors{Dunn, et al.}
\shorttitle{Evolution of the UV Outflows in the Seyfet 1 Galaxy NGC3516}

\title{Evolution of the Outflows in NGC 3516}

\author{Jay P. Dunn\altaffilmark{1}, 
Rozhin Parvaresh\altaffilmark{1},
S. B. Kraemer\altaffilmark{2}
\& D. Michael Crenshaw\altaffilmark{1} 
}

\altaffiltext{1}{Department of Physics and Astronomy, Georgia State University, Atlanta, GA 30303, USA: jdunn@gpc.edu}
\altaffiltext{2}{Department of Astronomy, The Catholic University of America}

\begin{abstract}

We analyse the 2011 HST/COS spectrum of the Seyfert 1 galaxy NGC 3516, which 
demonstrates clear changes in one of the intrinsic absorption troughs 
(component 5), slight evidence of change in a second trough (component 6), 
and the appearance of a new absorption trough (component 9). We interpret 
both the changes and appearance of the new trough as bulk motion across the 
line-of-sight. The implied lower limit on the transverse velocity of 
component 5 is 360 km s$^{-1}$ comparing to the earlier 2001 HST/STIS spectrum,
while the lower limits for components 6 and 9 are 920 km s$^{-1}$ based on 
2009 {\it FUSE} data. Component 5 also exhibits a shift in velocity centroid. 
This is only the second known case of this behavior in a Seyfert galaxy. 
Due to the high quality of the HST/COS spectrum, we identify a previously 
undetected trough due to an excited state of \siII\ for component 1. In 
combination with the resonance trough of \siII\ and photoionization 
modelling, we directly determine the distance of the component 1 outflow 
to be 67.2 pc.

\end{abstract}


\section{Introduction}

Active galactic nuclei (AGN) are extremely luminous objects 
due to matter accreting onto a supermassive black 
hole (SMBH). Seyfert galaxies are AGN that are relatively nearby 
(z $\leq$ 0.15) and only moderately luminous 
($log L_{bol} ergs^{-1} s$ = 43-45) 
and are known to be highly variable in continuum flux 
\citep[e.g.,][and references therein]{2006PASP..118..572D} 
on both short and long 
timescales. The variability is likely due to changes in the mass 
inflow rate of the accretion disk which is constrained to a small 
volume. In approximately 50\% of Seyfert galaxies, we observe 
blueshifted absorption troughs due to outflowing material ejected 
from the AGN 
\citep[i.e., intrinsic absorption][and references therein]{2003ARA&A..41..117C}.
These mass outflows in AGN 
are potentially an important feedback mechanism that help regulate 
the SMBH mass and explain the coevolution of the SMBH and galactic 
bulge \citep{2005Natur.433..604D,2010MNRAS.401....7H}. Similar to the 
nuclear continuum, absorption troughs due to outflows tend to show 
variability in structure or number in Seyfert galaxies. In one 
survey of this phenomenon, \citet{2008AJ....136.1201D} found a lower limit 
for absorption line variability in the far UV of approximately 40\% 
of a sample of 72 Seyfert Galaxies.

While the energetics of mass outflows are beginning to become 
constrained by observations, the dynamics of these winds is still 
unclear \citep[e.g.,][]{2004ApJ...600..580G,2005Natur.433..604D,
2008ApJ...686..219M,2010MNRAS.408L..95K,2012ApJ...745L..34Z,
2012ARA&A..50..455F,2012MNRAS.425..605F,2014MNRAS.439..400Z,
2014MNRAS.437.2404N,2014MNRAS.444.2355C,2015MNRAS.448L..30C,
2015ARA&A..53..115K}.  
In order to determine which driving mechanisms (radiation pressure,
MHD acceleration, thermal flows)
play important roles in the ejection of the material, we require critical 
information about the outflows in general such as transverse velocities. 
The most direct technique to constrain the transverse velocity requires 
variations in the troughs as observed over multiple epochs where the gas 
presumably moves into or out of the line-of-sight and is uncorrelated
with continuum variations, suggesting that ionization changes are not
responsible for the absorption variability
\citep{2001ApJ...557...30K,2007ApJ...659..250C}.

It follows that Seyfert galaxies provide excellent testbeds for 
exploring dynamical scenarios due to their proximity and highly 
variable nature. NGC3516 ($z$=0.00884) is a well-studied Seyfert 
galaxy that exhibits intrinsic 
absorption. \citet{2002ApJ...577...98K} characterized the absorption in 
the Space Telescope Imaging Spectrograph (STIS) 2001 observation as 8 
distinctive kinematic components and used photoionization modeling 
to place distance limits on each component from the nucleus. In 
X-ray spectra, \citet{2012ApJ...747...71H} found that the absorption 
structure showed variability over the span of $\sim$5 years. 
\citet{2012ApJ...753...75C} provided a summary of the 
physical constraints on both UV and X-ray absorbers in NGC 3516. 
We explore in this work the absorption trough variability of the 
ultraviolet lines in NGC3516 as well as the benefits of detecting 
previously unknown troughs due to the significantly higher 
signal-to-noise in the UV spectra from the Cosmic Origins 
Spectrograph (COS) onboard the                             
{\it Hubble Space Telescope} ({\it HST}).

\section{Update on the UV Spectra of NGC~3516}

Using the HST/COS, we obtained 
new high resolution (R=20,000) spectra of NGC~3516 
($z$=0.00884; RA: 11 06 47.491; Dec: +72 34 06.89) on 2011 January 
22. We list the details of these spectra in Table 1. 
We download the processed spectra \citep{2015cos..rept....1F}
from 
the Mikulski Archive for Space Telescopes (MAST) website. 
As each spectrum covers a different wavelength range, we 
coadd the spectra for a similar epoch and average 
overlapping regions weighting by exposure time. The final
coadded spectrum has an signal-to-noise ratio of 6.2 near
1550 \AA, which is 
2.5$\times$ higher than the previous HST/STIS spectra.

\newpage
\begin{deluxetable}{lcccc}
\tablecolumns{5}
\tablewidth{0.85\textwidth}
\tablecaption{New COS Observations of NGC~3516}
\tablehead{
\colhead{Dataset} &
\colhead{Start Time} &
\colhead{Exposure Time} &
\colhead{Grating} &
\colhead{Cent. Wavelength} \\
\colhead{} &
\colhead{(UT)} &
\colhead{(s)} &
\colhead{} &
\colhead{(\AA)} \\
}

\startdata
LBGU58010 & 06:19:33 & 560.032 & G130M & 1291  \\
LBGU58020 & 06:32:29 & 560.032 & G130M & 1300  \\
LBGU58030 & 06:45:25 & 560.000 & G130M & 1309  \\
LBGU58040 & 06:58:21 & 560.032 & G130M & 1318  \\
LBGU58050 & 07:11:59 & 684.192 & G160M & 1589  \\
LBGU58060 & 07:51:16 & 684.192 & G160M & 1600  \\
LBGU58070 & 08:06:07 & 684.192 & G160M & 1611  \\
LBGU58080 & 08:20:58 & 684.192 & G160M & 1623  \\

\enddata
\normalsize

\end{deluxetable}

We also obtained a 16,700 second spectrum of NGC~3516 on 
2007 January 23 with the {\it Far Ultraviolet Spectroscopic 
Explorer} \citep[{\it FUSE},][]{2000ApJ...538L...1M}. We download 
all of the available {\it FUSE} spectra of NGC~3516 from the MAST 
archive and process the spectra using the CalFUSE data 
reduction software \citep[v3.2.3,][]{2007PASP..119..527D}. To correct 
for the so-called ``worm'' feature frequently observed in the LiF2A 
segment, we fit the average ratio of the LiF2A to LiF1B with 
a spline and correct the LiF2A continuum to match the LiF1B 
\citep[similar to][]{2011A&A...534A..41K}. We coadd the 8 spectral 
{\it FUSE} segments and scale the fluxes in the overlapping 
regions of each segment to match the LiF1A, which is the 
primary channel used for pointing the instrument 
\citep[see][]{2008AJ....136.1201D}.

\newpage
\begin{deluxetable}{lcccc}
\tablecolumns{5}
\tablewidth{0.85\textwidth}
\tablecaption{FUSE Observations of NGC~3516}
\tablehead{
\colhead{Dataset} &
\colhead{Start Date} &
\colhead{Start Time} &
\colhead{Exposure Time} &
\colhead{Aperture} \\
\colhead{} &
\colhead{} &
\colhead{(UT)} &
\colhead{(s)} &
\colhead{} \\
}

\startdata
P1110404000 & 2000-04-17 & 10:58:04 & 16335 & LWRS \\
P2110102000 & 2002-02-14 & 01:46:54 & 20686 & LWRS \\
P2110103000 & 2003-01-28 & 06:17:55 & 16901 & LWRS \\
P2110104000 & 2003-03-29 & 05:44:39 & 16329 & LWRS \\
G9170101000 & 2006-02-09 & 17:12:24 & 28687 & LWRS \\
G9170102000 & 2007-01-23 & 06:38:51 & 16847 & LWRS \\
\enddata
\normalsize

\end{deluxetable}

\section{Changes in the Continuum Flux and Absorption Troughs}

\begin{figure}[!h]
  \centering \includegraphics[angle=90,width=1.0\textwidth]
  {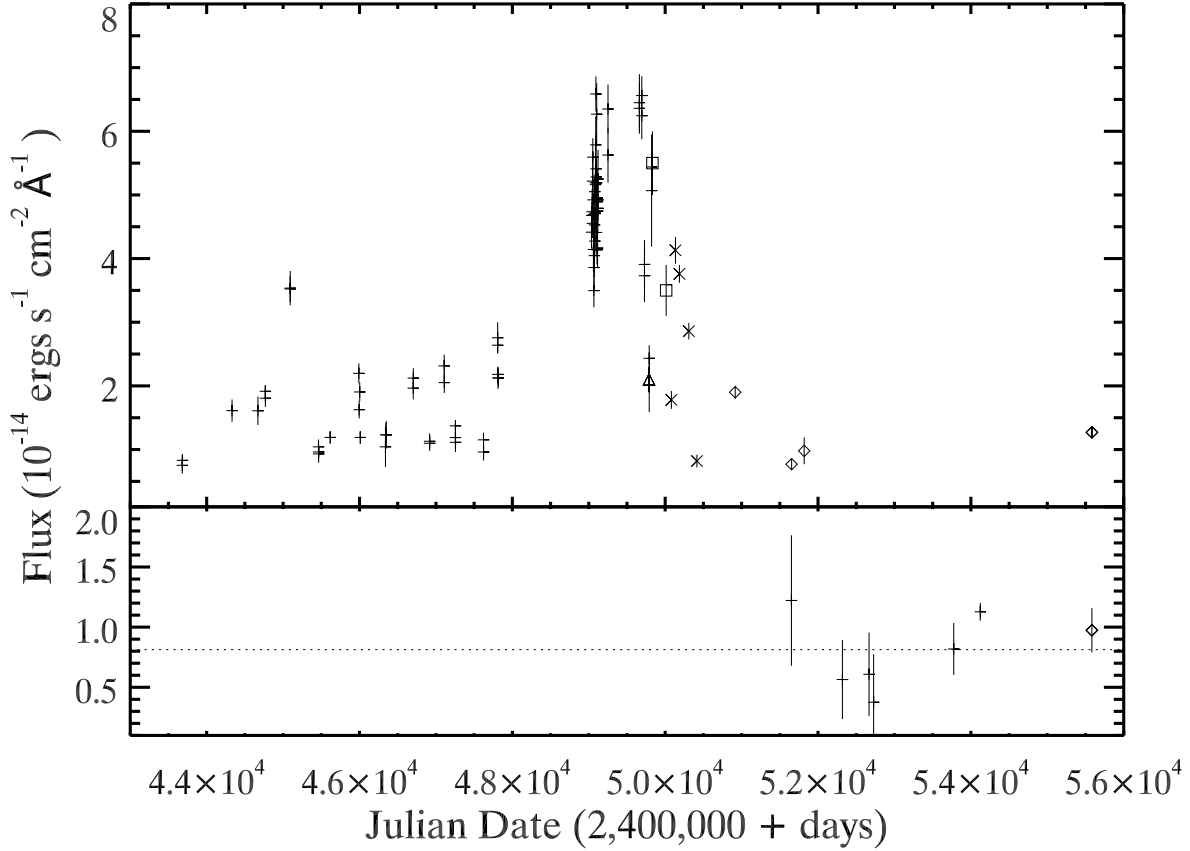}\\
  \caption[]
  {\footnotesize Top: UV continuum light curve of NGC3516. Average
flux values (in units of ergs s$^{-1}$ cm$^{-2}$ \AA$^{-1}$) are 
measured at 1460 \AA\ and plotted as a function of Julian date. 
The plus signs are fluxes taken from {\it IUE} observations, triangles from 
{\it HUT} observations, squares from GHRS observations, and crosses 
are from {\it HST} FOS spectra, and the diamonds are {\it HST} STIS 
observations. The most recent diamond is the average of the 2011 HST/COS 
spectra. The vertical lines indicate the uncertainties ($\pm$ 1$\sigma$). 
Bottom: FUV continuum light curve of NGC3516 measured at 1160 \AA. 
Crosses represent fluxes measured in FUSE spectra, while the 
diamonds are the fluxes measured from the 2011 HST/COS spectrum. The
horizontal dashed line represents the average flux level of the
measured data.}
  \label{f1}
\end{figure}

In Figure \ref{f1} (top), we show an updated UV continuum flux 
history of NGC~3516 measured at 1460 \AA\ in the observed frame
\citep{2006PASP..118..572D}. 
The continuum flux level in the HST/COS spectrum is consistent with 
the previous continuum flux measurements taken from HST/STIS spectra 
in 2000. This hints that the continuum flux of NGC~3516 has 
remained in a relatively low flux state over the last $\sim$10 
years. We also show the continuum flux lightcurve measured at 1160
\AA\ for the {\it FUSE} data in Figure \ref{f1} (bottom). While the
source has variation, the general flux level appears to have remained 
in a low state as seen in both the HST/COS and {\it FUSE} flux
measurement.

\begin{figure}[!h]
  \centering \includegraphics[angle=90,width=0.8\textwidth]
  {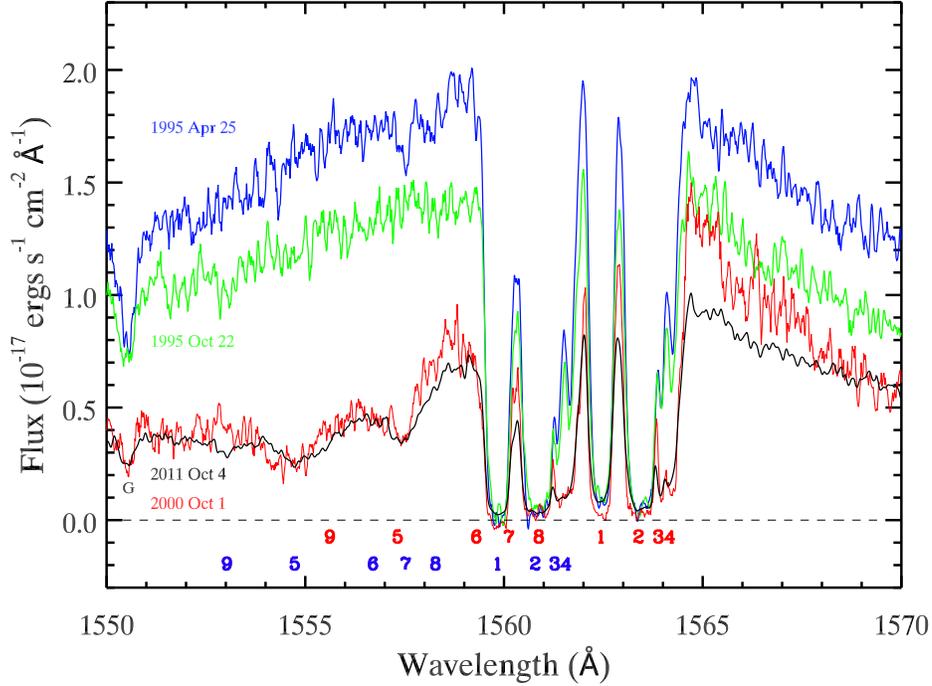}\\
  \caption[]
  {\footnotesize Plot of high resolution spectral segments 
showing the history of the \civ\ $\lambda\lambda$1548, 1551
emission line region with visible instrinsic absorption troughs 
in NGC3516. The red histogram is the 2000 spectrum taken by
HST/STIS, the black data are the 2011 spectrum taken by HST/COS, while the
blue and green histograms are data taken by GHRS in 1995 in April
and October, respectively. Both the GHRS and HST/STIS spectra are
boxcar smoothed by 5, while the HST/COS spectrum is not smoothed. 
We label both the previously identified troughs 
by Kraemer et al. (2002) as 1 through 8 (components 6 and 8 are 
only detected in \nv) as well as a newly detected component 9 trough.
The red and blue numbers mark the positions for the $\lambda$1548 and
$\lambda$1551 lines, respectively. We mark the position of the 
galactic \civ\ line at 1551 \AA. The horizontal, dashed line 
represents 0.0 flux.} 
  \label{f2}
\end{figure}

To evaluate the changes in the absorption troughs, we begin by 
plotting the 3 previous {\it HST} high-resolution spectra with 
the coadded HST/COS spectrum for 
\civ\ (Figure \ref{f2}) and identifying the previously known 
troughs \citep{2002ApJ...577...98K}. The most notable differences 
occur at high velocity. First, we label a new high velocity kinematic 
component 9 which is visible at approximately 1553 \AA\ or a 
radial velocity of $-$1700 km s$^{-1}$. This trough is detected at 
5$\sigma$ above the level of the flux uncertainties for the \nv\
doublet and 4 $\sigma$ for the \civ\ doublet. Secondly, the component 
5 trough visible in the HST/STIS spectrum, located at approximately 
1555 \AA, has had considerable change in the radial velocity 
of the centroid and overall profile. This difference in the 
radial velocity does not appear to be due to variation 
in the core emission profile 
as the change is evident after normalizing the emission.
Moreover, the component 7 trough, which is similar in depth 
and equivalent width and
closer to the emission core in radial velocity, does not exhibit 
any measurable change. Finally, we find that the 
component 1 trough in the red member of the \civ\ doublet 
($\lambda$1551) has become notably shallower.

\begin{figure}
\centering \includegraphics[angle=0,width=0.8\textwidth]
{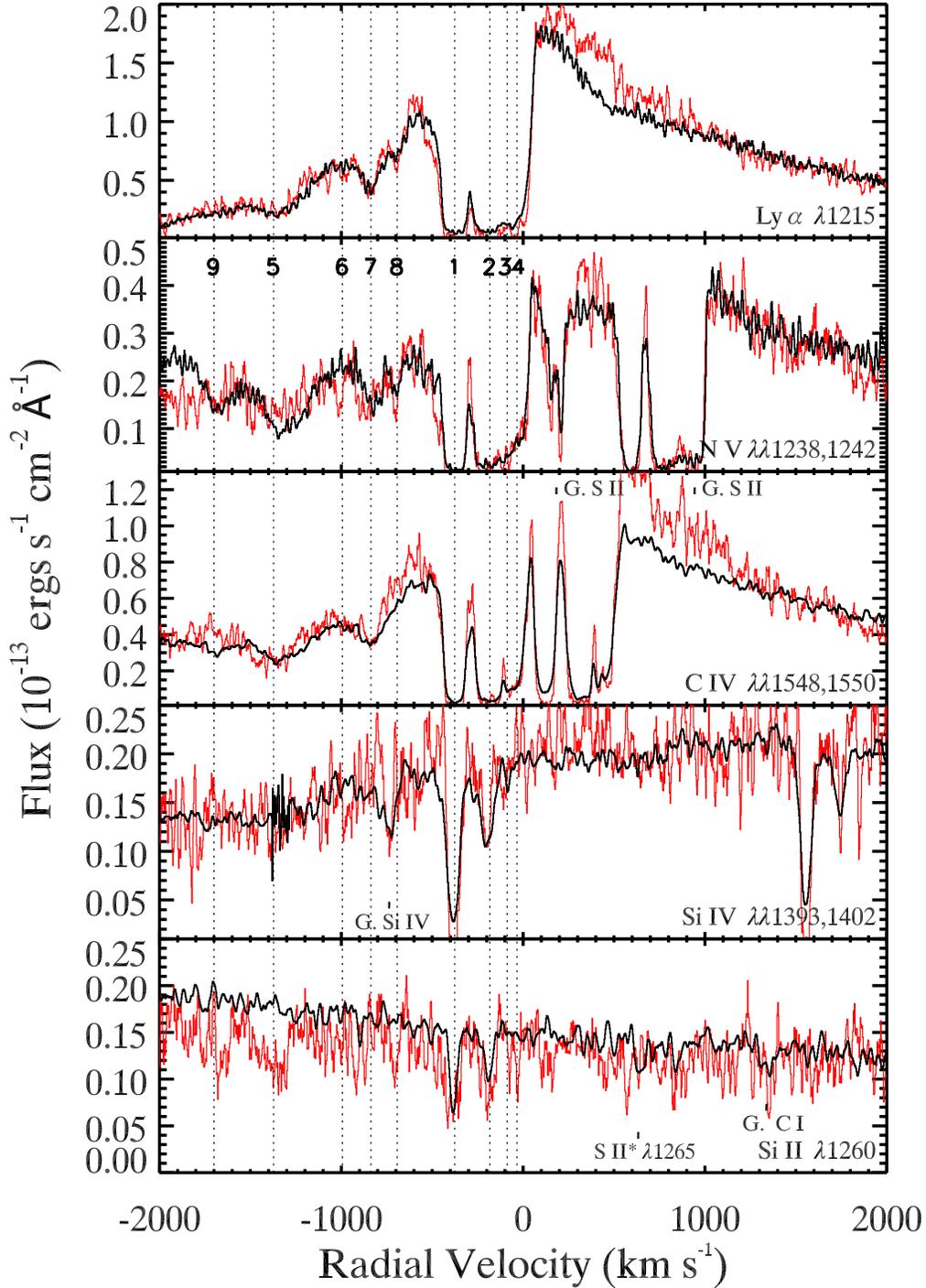}\\
\caption[]
{\footnotesize Plots of the HST/COS spectral segments (black) and
HST/STIS (red) for multiple ions in NGC3516 that show intrinsic absorption 
troughs. We plot the fluxes as a function of radial velocity. For doublet 
lines, we determine the velocities with respect to the blue components 
of the doublets (\nv, \civ, and \siiv). We place vertical dotted lines
at the positions of identified kinematic components for the blue
doublet members. For the HST/STIS spectra, we boxcar smooth the data by a 
factor of 5. We boxcar smooth the \siII\ plot by a factor of 9 to 
facilitate the clarity of the excited state \siII* $\lambda$1265 trough at
$\sim$600 km s$^{-1}$. Component 9 was not 
previously visible in spectra and
appears strongest in the \nv\ and \civ\ lines. Component 9 
for \Lya\ is very close to the galactic \Lya\ contamination, which 
is potentially obscuring the signal.}
  \label{f3}
\end{figure}

We plot in Figure \ref{f3} spectral segments from HST/STIS and the 
HST/COS for lines from commonly observed ions to search for changes in 
other components. Component 9 is clearly detected in both the 
\civ\ and \nv\ data but is not visible in \siiv\ or \Lya. The lack 
of a \Lya\ trough is likely due to the presence of the
Galactic \Lya\ trough. Similar to \civ, the component 5 
trough also appears to have undergone change in \nv, although 
the difference is not as distinct. Also noteworthy in the \nv\ 
doublet is the lack of the component 6 trough. The trough was 
only weakly detected by \citet{2002ApJ...577...98K} in 
the HST/STIS spectrum and has potentially vanished by the time of the 
HST/COS observation. This is based though on the poorer signal-to-noise 
level in the HST/STIS spectrum.
Finally, as with the \civ\ doublet, the component 1 trough in 
\siiv\ ($\lambda$ 1402) appears to be
significantly shallower in the HST/COS spectrum.

\begin{figure}
\centering \includegraphics[angle=0,width=0.8\textwidth]
{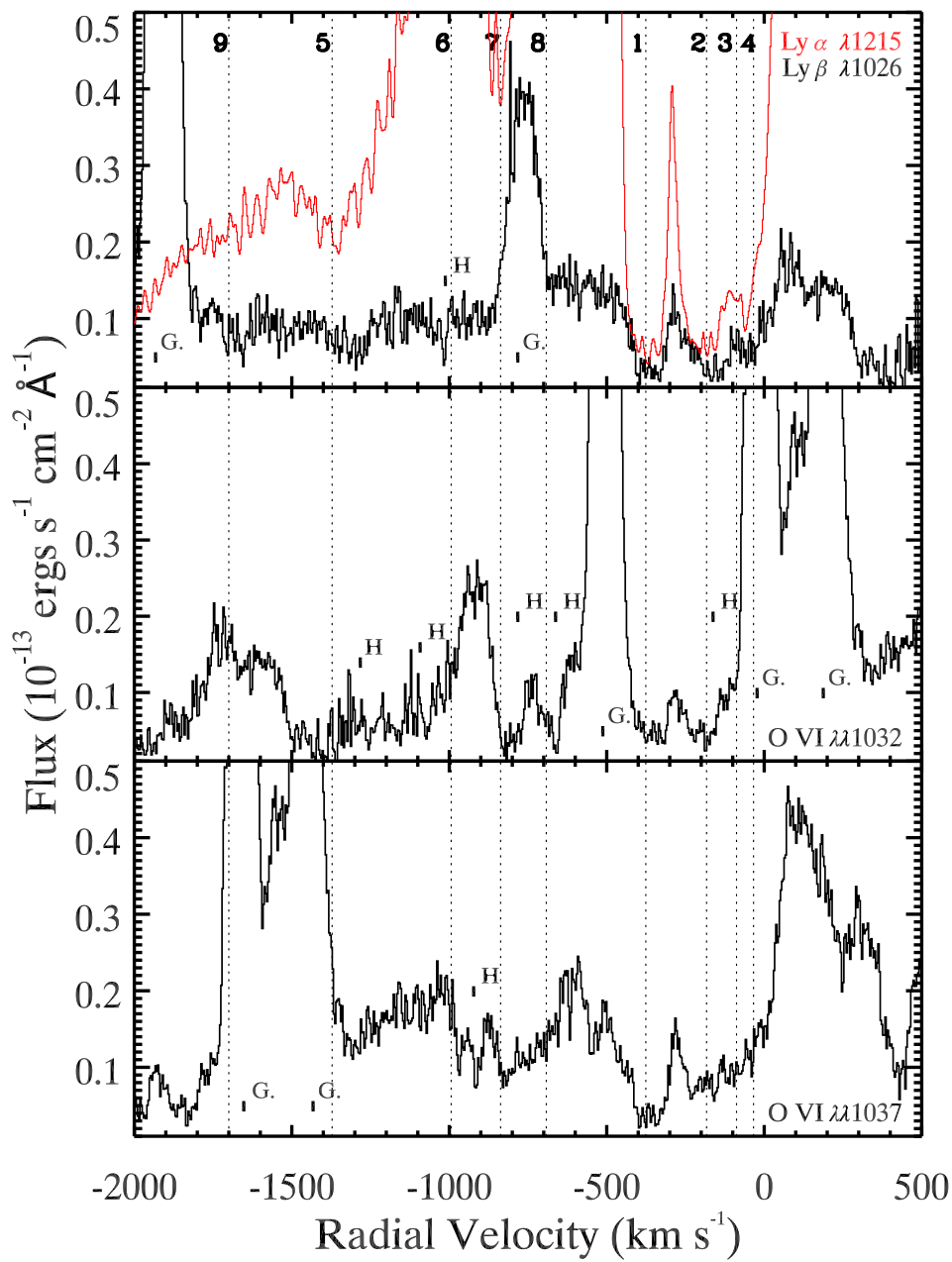}\\
\caption[]
{\footnotesize Plots of {\it FUSE} spectral segments for 
\Lyb (top) and the \ovi\ doublet $\lambda$1032 (middle),
$\lambda$1037 (bottom) similar to Figure 3. For \Lyb, we 
also plot the HST/COS spectrum of the \Lya\ region in red. Geocoronal
emission lines are marked with 'G.' and potential troughs
for the stronger H$_2$ lines due to the interstellar medium 
are marked with 'H'. It appears that component 9 is not visible 
for either ion and component 6 appears to have disappeared 
between the HST/STIS and {\it FUSE}  epochs.}
  \label{f4}
\end{figure}

We also plot in Figure \ref{f4} the {\it FUSE} spectra for 
\Lyb\ $\lambda$1026 and the
\ovi\ doublet $\lambda\lambda$1032,1037. We coadd the available
spectra to achieve the highest possible S/N ratio.
Component 9 does not seem to appear between the HST/STIS and {\it FUSE}
observations as it is not readily detected in either the \Lyb\ nor 
the \ovi\ $\lambda$1032 troughs, although it would be obscured by 
geocoronal emission in \ovi\ $\lambda$1037. Component 6 is also not 
detected in either ion. However, given it was not strongly detected 
in the HST/STIS spectrum it may not have strong associated \Lyb\ or 
\ovi\ troughs. Component 5 appears shallow in \Lyb\ and relatively 
correlated in velocity with the HST/COS data, which implies the 
change occurred between the {\it FUSE} and HST/COS observations.

\clearpage

\clearpage

We also detect the presence of a previously unobserved trough
due to the metastable, excited state line \siII* $\lambda$1265
for component 1. The trough is weak but clearly detected and
was probably not visible in the HST/STIS spectrum due to the
comparably lower signal-to-noise. The trough in conjunction 
with that from the resonance line \siII\ $\lambda$1260 seen in
both the HST/COS and HST/STIS data provides key
information on the number density of the gas and its distance
from the nucleus ($\S$5).

\section{Column Density Measurements}

\begin{figure}[!h]
  \centering \includegraphics[angle=0,width=0.8\textwidth]
  {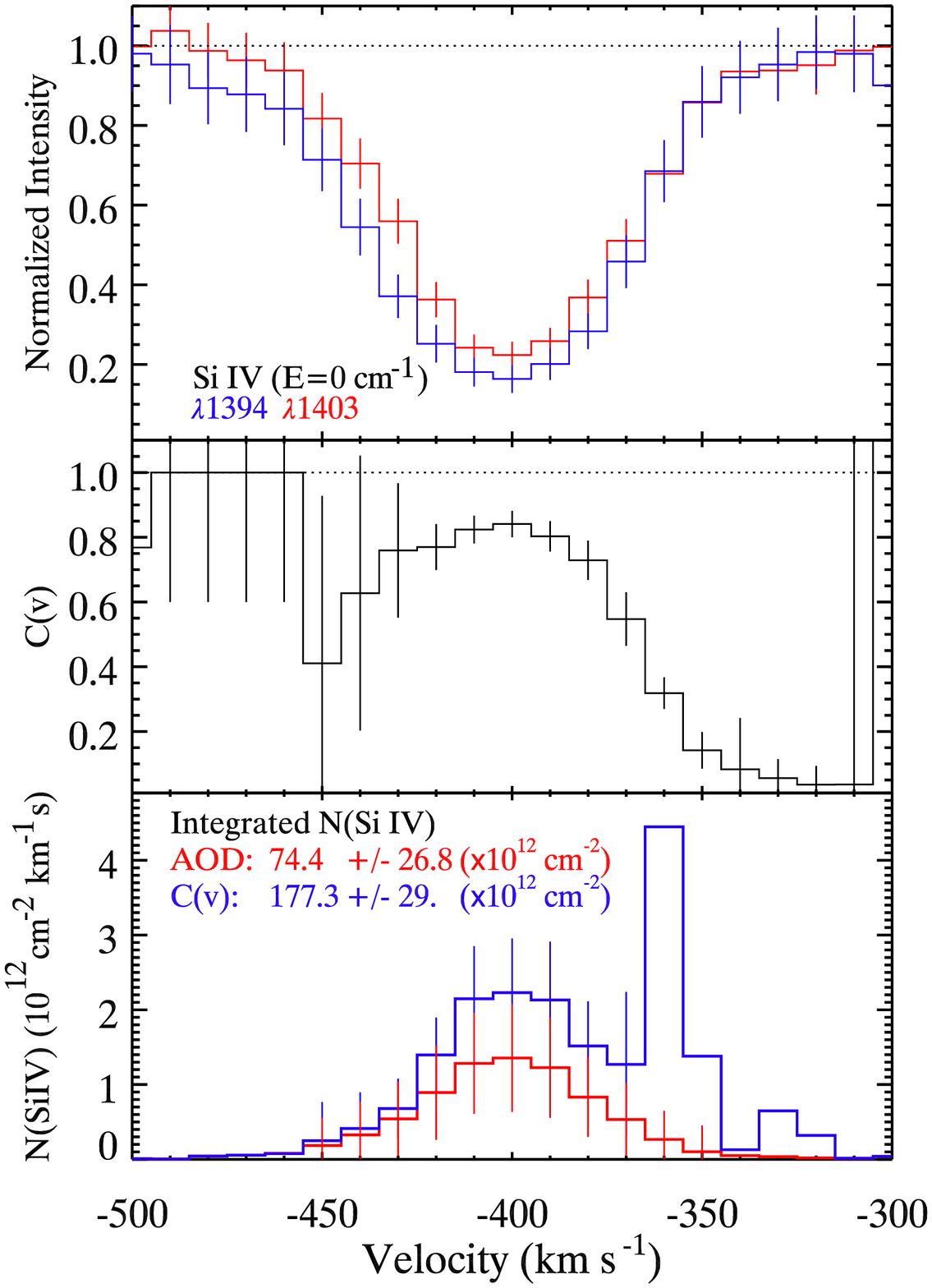}\\
  \caption[]
  {\footnotesize Top: Portions of the normalized HST/COS spectrum 
centered on the absorption troughs from \siiv $\lambda$1394 
(blue histogram) and $\lambda$1403 (red histogram) of kinematic 
component 1 plotted as a function of velocity. Vertical lines 
represent the statistical uncertainties in the residual 
intensity. 
Middle: Plot of the velocity-dependent covering factor ($C(v)$)
as determined from the doublet method as a function of velocity. 
We artificially set the value to a 
covering of 1.0 for unphysical points in the wings due to the
overlap of the blue and red lines from noise. For the unphysical
points near $-$350 km s$^{-1}$, we use the blue line trough depth
to determine the limit of $C(v)$.
Bottom: Velocity-dependent column density 
determinations for the \siiv\ lines. The lower red histograms 
show the best fit to the data based on the apparent optical
depth (AOD) while the upper blue histograms show 
the best fit from determining C($v$). The velocity-integrated 
\siiv\ column density values and their associated statistical 
errors are listed for each method.}
  \label{f5}
\end{figure}

To extract the ionic column densities ($N$) of the identified 
troughs, where available we use the velocity dependent pure 
partial covering method 
\citep[$C(v)$;][]{2002ApJ...580...54D,2005ApJ...620..665A,
2008ApJ...681..954A,2010ApJ...709..611D}. We 
simultaneously determine the optical 
depth and partial covering factor at every velocity element 
for each doublet or line series (i.e., the Lyman series), which 
have known ratios of oscillator strength. We show an example 
of this for the column density of \siiv\ in Figure $\ref{f5}$. 
For \siII\ and \cii\ lines of component 1, we assume the 
covering factor from their higher ionization state counterparts 
(i.e., \civ\ and \siiv). For single lines or to 
provide an absolute lower limit for the column density, we 
also calculate the apparent optical depth 
\citep[AOD;][]{1991ApJ...379..245S}. We also measure the new 
column density of 
components 5 and 9 based purely on AOD for the blue members
of the doublets as the red are blended with other components. 
We list in Table 3 the new column density determinations for 
components 1, 5, and 9 and the previously measured column
density values for components 1 and 5 from 
\citet{2002ApJ...577...98K} for
comparison. For component 1, the measured column densities
have exhibited little to no change between epochs, despite
the slight change in the depths of some of the troughs.

\newpage
\begin{deluxetable}{lccccc}
\tablecolumns{6}
\tablewidth{0.72\textwidth}
\tablecaption{Ionic Column Densities$^a$}
\tablehead{
\colhead{Ion} &
\colhead{N$_{ion}$(C1)} &
\colhead{N$_{ion}$(C1)} &
\colhead{N$_{ion}$(C5)} &
\colhead{N$_{ion}$(C5)} &
\colhead{N$_{ion}$(C9)} \\
\colhead{} &
\colhead{(COS)} &
\colhead{(STIS)$^b$} &
\colhead{(COS)} &
\colhead{(STIS)$^b$} &
\colhead{(COS)} \\
}

\startdata
\cii    & 73.2 $\pm$ 10.6 & 79 $\pm$ 25  & $-$    & $<$25        & $-$   \\
\cii *  & 99.1 $\pm$ 11.6 & 144 $\pm$ 30 & $-$    & $<$25        & $-$   \\
\ciii   & $<$14           & $-$          & $-$    & $-$          & $-$   \\
\civ    & $>$495          & $>$1026      & $>$160 & 211 $\pm$ 30 & $>$50 \\
\siII   & 15.5 $\pm$ 1.4  & $-$          & $-$    & $-$          & $-$   \\
\siII * & 4.7 $\pm$ 1.3   & $-$          & $-$    & $-$          & $-$   \\
\siiv   & 126 $\pm$ 51    & $>$186       & $-$    & $<$4         & $-$   \\
\hi     & $>$980          & $>$599       & $-$    & 97 $\pm$ 18  & $-$   \\
\ovi    & $>$1020         & $-$          & $-$    & $-$          & $-$   \\
\nv     & $>$1090         & $>$2048      &        & 165 $\pm$ 39 & $>$150\\
\pv     & $<$12           & $-$          & $-$    & $-$          & $-$   \\
\enddata

\normalsize
\tablenotetext{a}{Units of 10$^{12}$ cm$^{-2}$.}
\tablenotetext{b}{As measured by Kraemer et al. (2002).}

\end{deluxetable}

For the hydrogen column density, we use the higher order Lyman 
lines visible in the $\it FUSE$ spectra as there appears to be
little evidence for change in continuum flux. The $\Lyb$ 
$\lambda$1026 trough appears to be shallower than the $\Lya$
$\lambda$ 1216 trough in a normalized spectrum, which suggests 
non-saturated troughs. However, the underlying continuum in the 
$\it FUSE$ spectrum 
is probably contaminated due to the much larger aperture. 
Thus, we treat the measured column density from $\Lyb$ as 
a lower limit. It is also noteworthy that in the 
$\Lyg$ $\lambda$973 trough, despite the poor signal-to-noise, 
the component 1 trough clearly appears to have a narrow high 
velocity sub-component and hints of a second lower velocity 
trough.

\begin{figure}
\centering \includegraphics[angle=90,width=1.0\textwidth]
{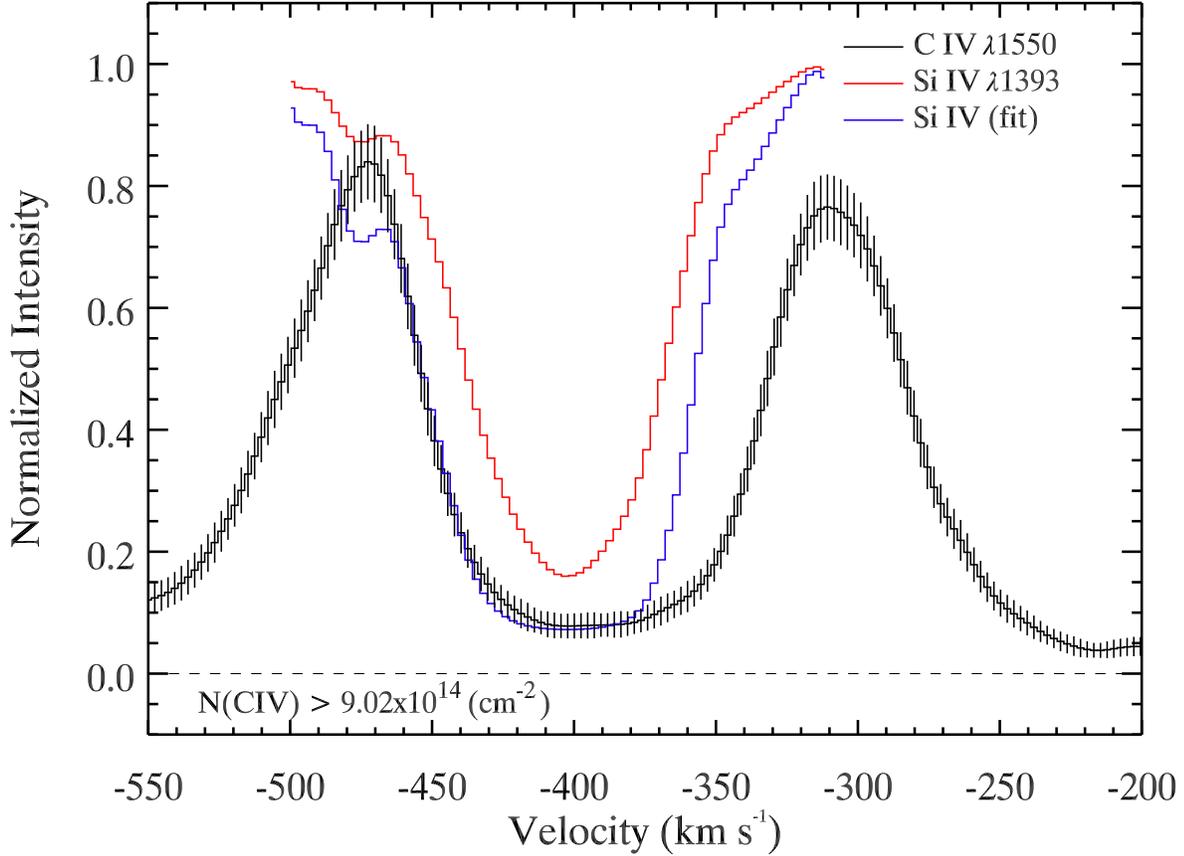}\\
\caption[]
{\footnotesize Plot of normalized flux as a function of
velocity of the spectral region for \civ\ $\lambda$1551 
(in black) showing component 1. We also plot the \siiv\
$\lambda$1393 trough for comparison. Using the \siiv\
trough as a template, we scale the trough in optical 
depth until the trough wall matches the \civ\ wall, 
shown in blue. While
the blue wing of the saturated template does agree with 
the \civ\ trough, the red wing is radically disparate. 
Unless the trough shape is entirely determined by 
partial covering, the \siiv\ template cannot adequately
match the \civ\ trough. To determine 
the column density limit for \civ\ over the same range 
as the \siiv\ trough, we integrate the scaled optical 
depth of the \siiv\ template.}
  \label{f6}
\end{figure}

The $C(v)$ and AOD methods only provide lower limits for 
traditionally saturated troughs such as the \civ\ doublet 
$\lambda\lambda$1548,1551. To attempt a more accurate 
determination, we begin by creating a template from the 
\siiv\ $\lambda$1402 trough, which appears to be 
non-saturated as it has a notably shallower depth than its 
doublet counterpart ($\lambda$1393). We scale the template 
in optical depth until the wing of the template matches the 
trough wall (see ``wing'' fitting technique in Dunn et al. 
2010). We show the fit for the \civ\ $\lambda$1551 line
in Figure $\ref{f6}$, which provides the largest lower
limit on the column density. From the match, it is 
clear that the 
\siiv\ template cannot match the full velocity range of 
the \civ\ troughs when we scale the trough to match the wing,
which we discuss further in the next section.

\clearpage

Finally, we find the upper limits for column densities of 
\ciii\ and \pv\ for component 1. Because there are no clear 
indications of troughs
for either the \pv\ doublet $\lambda,\lambda$1118, 1128 or 
the \ciii* excited state line $\lambda$1175, we scale
the \siiv\ template for component 1 and compare it to the
region where these lines would be detected. We vary the scaling 
in optical depth to the noise level to find the largest possible 
optical depth that could be buried without detection. We 
integrate across the template to determine the column density
limits and list the values in Table 2.




\section{Implications of the New Column Densities and Variability}

We find four kinematic components in absorption that show evidence 
of change despite very little evidence for significant changes 
in the continuum 
flux. Comparing the measured column densities for
component 1 in the 
HST/COS spectrum to those measured by \citet{2002ApJ...577...98K} for the 
HST/STIS spectrum (also listed in Table 3), we find that the column 
densities do not appear to 
have changed. Thus, the photoionization model determined in 
\citet{2002ApJ...577...98K} is likely still viable for the HST/COS 
data and 
further supports the argument that NGC~3516 has been stable since 
the 2000 HST/STIS observation.

Because of the apparent lack of change in flux, column densities, 
and ionization state, both 
the appearance of component 9, the disappearence of component 6
and the changes in component 5 are 
most easily explained as material moving into or out of the 
line-of-sight due to transverse velocity. As these are absorbing
light from the  broad line region (BLR), we limit the transverse
velocity based on the crossing time across the line-of-site to 
the BLR. The size of the BLR in \civ\ for NGC3516 is estimated to be
approximately 4.5 ld \citep{1999ApJ...524..707G} and the time
between the last HST/STIS and HST/COS observations is 4020 days. 
This yields a lower limit on the transverse velocities for
components 5, 6, and 9 of greater than 360 km s$^{-1}$. However, since
the troughs for components 6 and 9 were also not detectable in the 
$\it FUSE$ spectrum of
2007, we estimate a more robust lower limit of 920 km s$^{-1}$.

Examining the \civ\ $\lambda$1551 trough of component 1, the 
slight change could also be an indication of a changing column 
density. Unlike components 5 and 9, since the change only 
occurred in \civ, it is more plausible that 
the difference between the HST/STIS and HST/COS spectra is due to the 
larger aperture of HST/COS. The HST/COS spectrum likely contains an 
underlying continuum and emission due to active star formation. 
The structure of component 1 as a whole appears to be a blend
of two distinct kinematic components as evidenced by the fit
to \civ\ and the \Lyg\ trough. This potentially provides
observational support to the two photoionization models 
necessary (1a and 1b) to match the ionic column densities 
measured for component 1 \citep{2002ApJ...577...98K}.

While the column densities we measure are similar to those of
\citet{2002ApJ...577...98K}, due to the higher signal-to-noise we
detect and measure the column density of an excited state line 
of \siII* for 
component 1. Assuming the plasma is in equilibrium,
the ratio of the \siII* column density to the column density
of the \siII\ resonance line at $\lambda$1260 provides
a simple and reliable diagnostic of the electron density 
\citep{2009A&A...508.1527B}. From the measured column densities,
we determine a density of $log n_H/cm^{-3}\approx$2.3 for component 1. 

The ionization state of the plasma is described by the ratio
of the rate of hydrogen ionization to recombination, or the 
so-called ionization parameter given by: 
\begin{equation}
U_H = \frac{Q_H}{4 \pi R^2 n_H c},
\end{equation}
where $Q_H$ is the rate of hydrogen ionizing photons emitted
by the AGN, $R$ is the radial distance of the plasma from the
nucleus, and $c$ is the speed of light. We measure $Q_H$ from
the HST/COS spectrum assuming a ``UV-soft'' SED (Dunn et al. 2010) 
and find $log(Q_H)$=53.6. Using the value of $log(U_H)$ of 
$-$0.9 from \citet{2002ApJ...577...98K} for component 1a, we find 
a distance to the absorber from the nucleus of 67.2 pc. 
This distance is consistent with the lower limit derived by 
\citet{2012ApJ...753...75C} of 15.5 pc.

\section{Conclusions and Discussion}

Using HST/COS spectra of NGC3516, we find for the first time in NGC3516
the appearance of a new absorption
trough (component 9) between the FUSE observation in 
2006 and the HST/COS observation in 2012. Previous changes in troughs
were all ascribed to ionization changes (Kraemer et al. 2002). 
We also show that component
5 has undergone measurable change in the trough structure and
radial velocity, which we attribute to bulk motion. The only 
other known case of radial velocity change is NGC3783 
\citep{2003ApJ...595..120G}. The final change we find in the 
new spectrum is that component 6, which was a weakly detected 
trough in the HST/STIS spectrum, is no longer visible. 
Accepting the lower signal-to-noise of the HST/STIS spectrum, this 
disappearance could be due to motion out of the line-of-sight.
Considering the physical properties of the AGN of NGC3516, 
these changes imply lower limits on the transverse 
velocity of component 5 of 360 km s$^{-1}$ and components 9 and 
potentially component 6 of 920 km s$^{-1}$. 

We also find a previously undetected trough due to the higher 
signal-to-noise HST/COS spectrum of an excited state line for \siII*. 
We measure the column density for \siII* and use the previously
established photoionization models by Kraemer et al. (2002) to 
determine the radial distance of component 1 to be 67.2 pc. 
Due to the radiation shielding arguments established in Kraemer 
et al. (2002), this distance agrees with the upper limits
for components 2, 3, and 4 but provides a new lower limit for
components 5$-$8 that is $\sim$4.3x larger than the previous
limit. Crenshaw \& Kraemer (2012) estimated the total kinetic 
luminosity output for NGC~3516 to be 
$\dot{K}$ $>$ 5.4$\times10^{41}$ ergs s$^{-1}$ based 
on the previous limits. 
Given the increased limit on distance and the linear relation
between distance and $\dot{K}$, the new lower limit for 
the total kinetic luminosity of NGC~3516 is 
$\dot{K}$ $>$ 2.4$\times10^{42}$ ergs s$^{-1}$.

Examining the transverse velocities of components 5 and 9
and their radial components of $-$1450 km s$^{-1}$ and 
$-$1700 km s$^{-1}$, respectively, we find that the limit on 
the transverse motion of component 9 (and perhaps 5) is 
comparable to the radial motion. Similar to the other higher
velocity components, component 9 exhibits a high column density
ratio for \nv/\civ, which implies either that it is highly ionized 
or that it's physical state is due to screening by component 1
(Kraemer et al. 2002). Because the other components with 
similar column density ratios only appear in low flux states, it
follows that they are likely experiencing screening and are at 
larger radial distances than 
component 1. The column densities for component 9 are consistent
with this scenario and suggest that it also has a radial distance
beyond that of component 1. Given the limit on the distance from 
the nucleus for both components 5 and 9, the transverse velocities 
for each component are larger than the rotational velocities, which
we estimate to be 67 km s$^{-1}$ based upon the SMBH mass 
\citep[42.7 million $M_\odot$;][]{2004ApJ...613..682P} and 
gravitational potential of the host galaxy 
\citep{1992ApJ...394...91M}. The large transverse
velocities beyond that of galactic rotation could suggest an origin 
much closer to the nucleus.


\bibliographystyle{apj}

\bibliography{ms}

\end{document}